\documentclass[aps,prl, twocolumn, amsmath, amssymb, floatfix, showpacs]{revtex4-1}
\usepackage[final]{graphicx}
\usepackage{hyperref}
\usepackage{subfigure}
\usepackage{enumerate}
\graphicspath{{Figures/}}

\begin{document}
\title{Effect of Interface Induced Exchange Fields on  Cuprate-Manganite Spin  Switches}

\author{Yaohua~Liu$^{1}$} \email[]{yhliu@anl.gov}
\author{C.~Visani$^{2}$} \altaffiliation[Present address:]{Unite Mixte de Physique CNRS Thales. Palaiseau, France}
\author{N.~M.~Nemes$^{2}$}
\author{M.~R.~Fitzsimmons$^{3}$}
\author{L.~Y.~Zhu$^{1}$}
\author{J.~Tornos$^{2}$}
\author{M.~Zhernenkov$^{3}$} \altaffiliation[Present address:]{Argonne National Laboratory, Argonne, Illinois 60439}
\author{A.~Hoffmann$^{1}$}
\author{C.~ Leon$^{2}$}
\author{J.~Santamaria$^{2}$}
\author{S.~G.~E.~te~Velthuis$^{1}$} \email[]{tevelthuis@anl.gov}

\affiliation{$^{1}$Materials Science Division, Argonne National Laboratory, Argonne, Illinois 60439, USA}
\affiliation{$^{2}$GFMC, Departamento de Fisica Aplicada III, Universidad Complutense de Madrid, Campus Moncloa, ES-28040 Madrid, Spain}
\affiliation{$^{3}$Los Alamos National Laboratory Los Alamos, New Mexico 87545, USA}

\date{\today}

\begin{abstract}

We examine the anomalous inverse spin switch behavior in La$_{0.7}$Ca$_{0.3}$MnO$_3$ (LCMO)/YBa$_2$Cu$_3$O$_{7-\delta}$ (YBCO)/LCMO trilayers by combined transport studies and polarized neutron reflectometry. Measuring magnetization profiles and magnetoresistance in an in-plane rotating magnetic field, we prove that, contrary to many accepted theoretical scenarios, the relative orientation between the two LCMO's magnetizations is not sufficient to determine the magnetoresistance. Rather the field dependence of magnetoresistance is explained by the interplay between the applied magnetic field and the (exponential tail of the) induced exchange field in YBCO, the latter originating from the electronic reconstruction at the LCMO/YBCO interfaces.

 \end{abstract}

\maketitle

Interfacial electronic reconstruction offers the possibility to engineer the electronic ground state with unprecedented access to exotic phenomena at epitaxial interfaces of complex oxide heterostructures, such as metallicity, superconductivity (SC) and even ferromagnetism (FM) at the interface of two insulating and non-magnetic oxides~\cite{OhtomoNature2004, ReyrenScience2007, BrinkmanNatMat2007}.  Another example is the interface between half-metallic ferromagnet (FM) La$_{0.7}$Ca$_{0.3}$MnO$_3$ (LCMO) and high $T_C$ superconductor (SC) YBa$_2$Cu$_3$O$_{7-\delta}$ (YBCO), where electronic reconstruction yields an anti-ferromagnetic coupling between the Cu and Mn's spins~\cite{SalafrancaPRL2010}, which generates an interface induced ferromagnetic exchange field on the Cu ions in YBCO.  This induced exchange field in YBCO then gives rise to a net Cu moment, as has been experimentally observed~\cite{ChakhalianNatPhy2006, WernerPRB2010, VisaniPRB2011}.

LCMO/YBCO/LCMO (LYL) trilayers are of interest as they are high-$T_C$ superconducting spin switches, yet exhibit the  so-called \textit{inverse} superconducting spin switch behavior. It has been shown that, in the superconducting transition region, LYL trilayers have lower resistances when the magnetizations of two ferromagetic layers are parallel, and higher resistances when they are antiparallel~\cite{PenaPRL2005}. Consequently, $T_C$ is higher for the parallel state and lower for the antiparallel state, which is opposite to the expectation based on the conventional proximity effect~\cite{TagirovPRL1999, BuzdinEPL1999}. The origin of the inverse spin switch behavior is still controversial. Possible mechanisms include the effect of stray fields~\cite{ZalkPRB2009, StamopoulosSST2007}, an imbalance of quasiparticles~\cite{TakahashPRL1999, NemesPRB2008}, and triplet superconductivity~\cite{DybkoPRB2009}. In these scenarios, the magnetoresistance depends on the relative magnetic alignment between the two ferromagnets, and the applied field direction only plays an indirect role by changing the magnetization configuration~\cite{TagirovPRL1999, BuzdinEPL1999, TakahashPRL1999, ZhuPRL2010}. Alternatively, Salafranca and Okamoto have recently proposed a scenario that can explain the inverse superconducting spin switch effect in LYL trilayers, in which the direction of the applied field plays a direct role. They argue that the superconductivity in the central YBCO is governed by the total field $\vec{H}_{tot}$ that results from the superposition of the applied field $\vec{H}_{a}$ and (the tail of) the aforementioned exchange field in YBCO, $\vec{H}_{ex}$, in a way similar to the magnetic field induced superconductivity~\cite{JaccarinoPRL1962, MeulPRL1984}. Accordingly, the alignment between $\vec{H}_{a}$ and $\vec{H}_{ex}$ influences the superconductivity in YBCO, and consequently a modulation in the alignment between $\vec{H}_{ex}$ and $\vec{H}_{a}$ should accompany a change of resistance in the superconducting transition region.

In this Letter we examine the angular dependence of the magnetization structures in LYL trilayers in experiments where the magnetic field rotates in-plane.  We utilize the polarized neutron reflectometry (PNR) technique, which is capable of resolving the depth profile of the magnetization with sub-nanometer resolution~\cite{FelcherRSI1987, MajkrzakPhysB1996, Fitz2005}, to correlate the angular dependent magnetization structure and magnetoresistance (MR).   We show unambiguously that,  in the superconducting transition region, MR  depends on the alignment between $\vec{H}_{ex}$  and $\vec{H}_{a}$, rather than the alignment between the two LCMO's magnetizations. This result strongly supports the Salafranca-Okamoto's scenario and settles a longstanding debate.

Samples were grown by sputter deposition in pure oxygen atmosphere on (100) SrTiO$_3$ substrates~\cite{SefriouiPRB2003} with a nominal structure of 40~unit-cells (u.c.) LCMO/8~u.c.YBCO/40~u.c. LCMO.   The sample size is $5 \times 10$~mm$^2$.  X-ray reflectometry (XRR) experiments were conducted at room temperature using Cu $K_{\alpha}$ radiation.  Polarized neutron reflectometry (PNR) experiments were conducted on the ASTERIX reflectometer at the Lujan Neutron Scattering Center. Magnetic hysteresis loops, magneto-transport data and PNR data were taken at 26~K . The sample's resistance is $\sim 10^{-4}$ of the normal state resistance at 26 K so that  the magnetoresistance is overwhelmed by the modulation of the superconductivity in the YBCO layer. 

\begin{figure}[tb]
	\centering
		\includegraphics[width=0.5\textwidth]{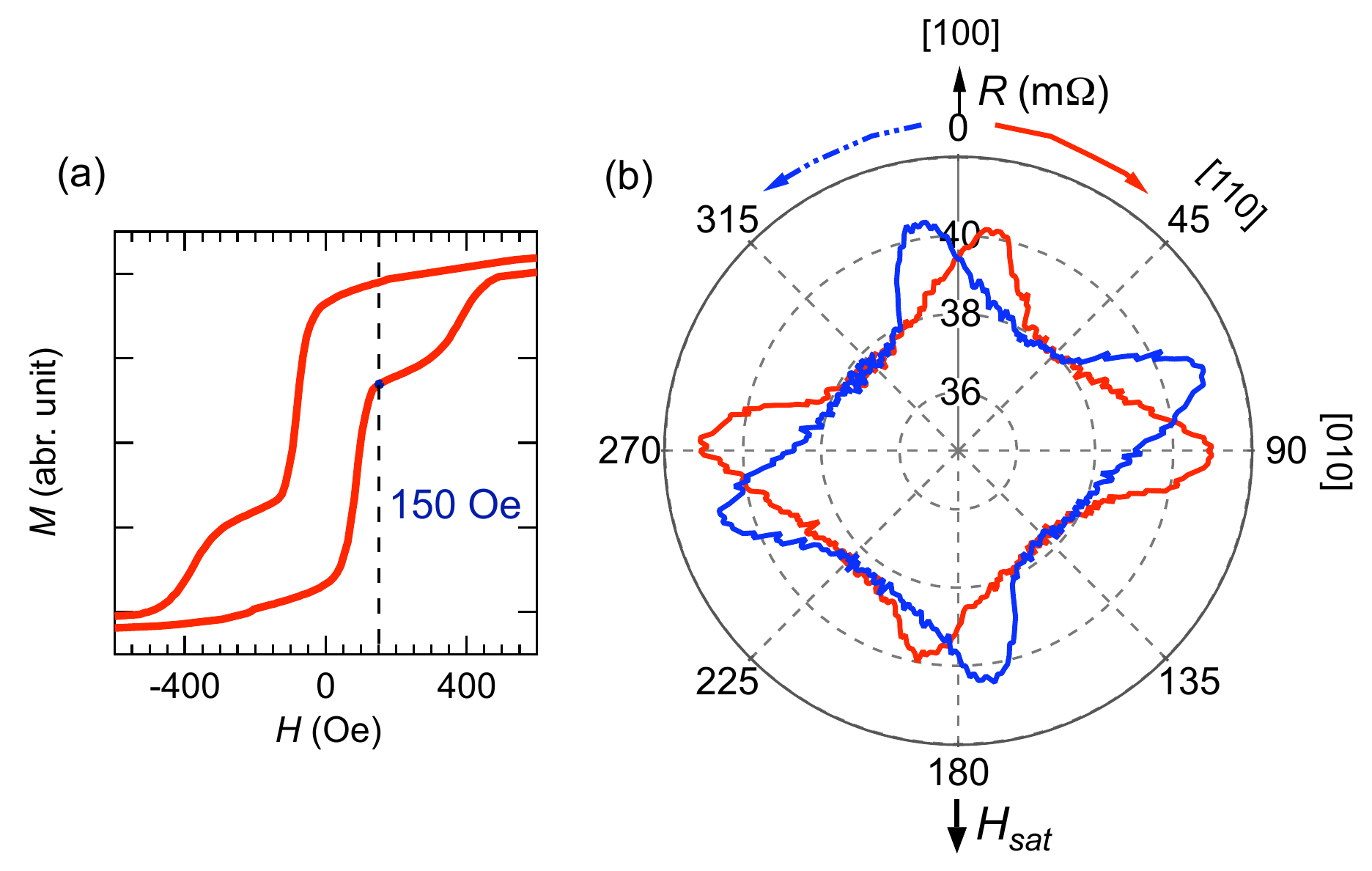} 
		\caption{\label{Fig:MR}(Color online) (a) Magnetization hysteresis loop along an easy axis ([110]). The dashed line shows $H = 150$~Oe. (b) Angular dependence of MR. A 150~Oe in-plane field is applied after having negatively saturated the film along 180$^\circ$ (a hard axis direction). Resistances are recorded when the field rotates either clockwise (CW) or counter-clockwise (CCW).}
\end{figure}

Our LCMO films have an in-plane cubic anisotropy with the easy axes along the $[110]$ and $[1\bar{1}0]$ axes~\cite{VisaniPRB2011}. Figure~\ref{Fig:MR}(a) shows the easy-axis magnetization hysteresis loop. Presumably due to different strain states of the bottom LCMO layer (grown on SrTiO$_3$) and the top LCMO layer (grown on YBCO), the two layers have different magnetic properties. The well-separated two-step switching with different step sizes indicates different saturation magnetizations, and different anisotropies between the top and bottom LCMO layers. Therefore, the relative magnetization orientation in the top and bottom layers is modulated upon rotating in an in-plane magnetic field, with an amplitude between the two coercivities. Figure~\ref{Fig:MR}(b) shows the magnetoresistance in a polar plot for a field of fixed magnitude (150 Oe).  The further from the radius origin the larger the resistance. The field direction $\Phi_{H}$ is defined  with respect to [100] direction. The MR shows a quasi-four-fold symmetry with four local $R_{min}$'s along the LCMO's magnetic easy-axis directions, \textit{i.e.}, 45$^\circ$, 135$^\circ$, 225$^\circ$ and 315$^\circ$; it  also shows a hysteresis between clockwise (CW) and counter-clockwise (CCW) rotations. 

\begin{figure}[tb]
	\centering
		\includegraphics[width=0.5\textwidth]{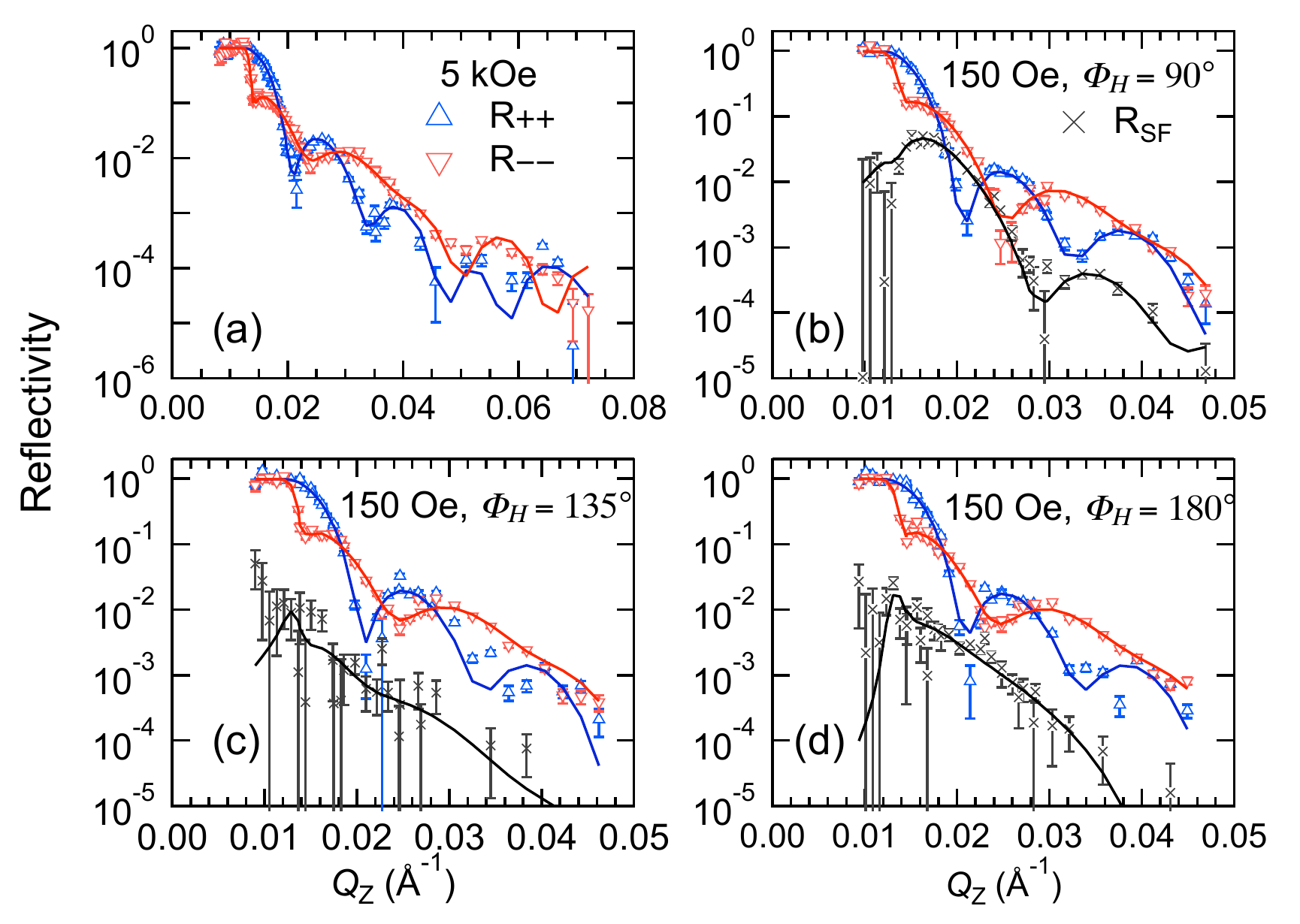}
	\caption{\label{Fig:PNR}(Color online) (a) PNR data at 5 kOe (the sample is in saturation). (b-d) Representative PNR data at 150~Oe with field direction $\Phi_H =$ (b) $90^\circ$,   (c) $135^\circ$ and (d) $180^\circ$, respectively. Symbols are the experimental data and the lines show the the best fits.} 
\end{figure}

We determine the saturated magnetizations of the top and bottom LCMO layers with complementary studies of XRR and PNR. Figure~\ref{Fig:PNR}(a) shows the PNR data in saturation with a 5~kOe field applied along the [100] direction. The reflectivities are plotted versus the wavevector transfer along the film's normal direction $Q_z$. $Q_z = 4 \pi sin(\theta_{i}) / \lambda$, where $\theta_{i}$ is the incident angle and $\lambda$ is the neutron's wavelength.  $R^{++}$ and $R^{--}$ are the two non-spin-flip reflectivities. With a combined refinement of the XRR and PNR data, we find that the saturation magnetizations of the top and bottom LCMO layers are 380 and 540~emu/cm$^3$, respectively. As also reported previously, the fitting indicates a possible suppression of the magnetization at the LCMO/YBCO interfaces~\cite{HoffmanPRB2005}. However, because of the limited $Q_z$ range, this PNR study is not sufficient to resolve the subtleness of the magnetization profile at the interfaces so that the amplitude of YBCO magnetization cannot be determined accurately. (See Supplemental Material~\cite{SM} for further details.)

 \begin{figure}[tb]
	\centering
		\includegraphics[width=0.5\textwidth]{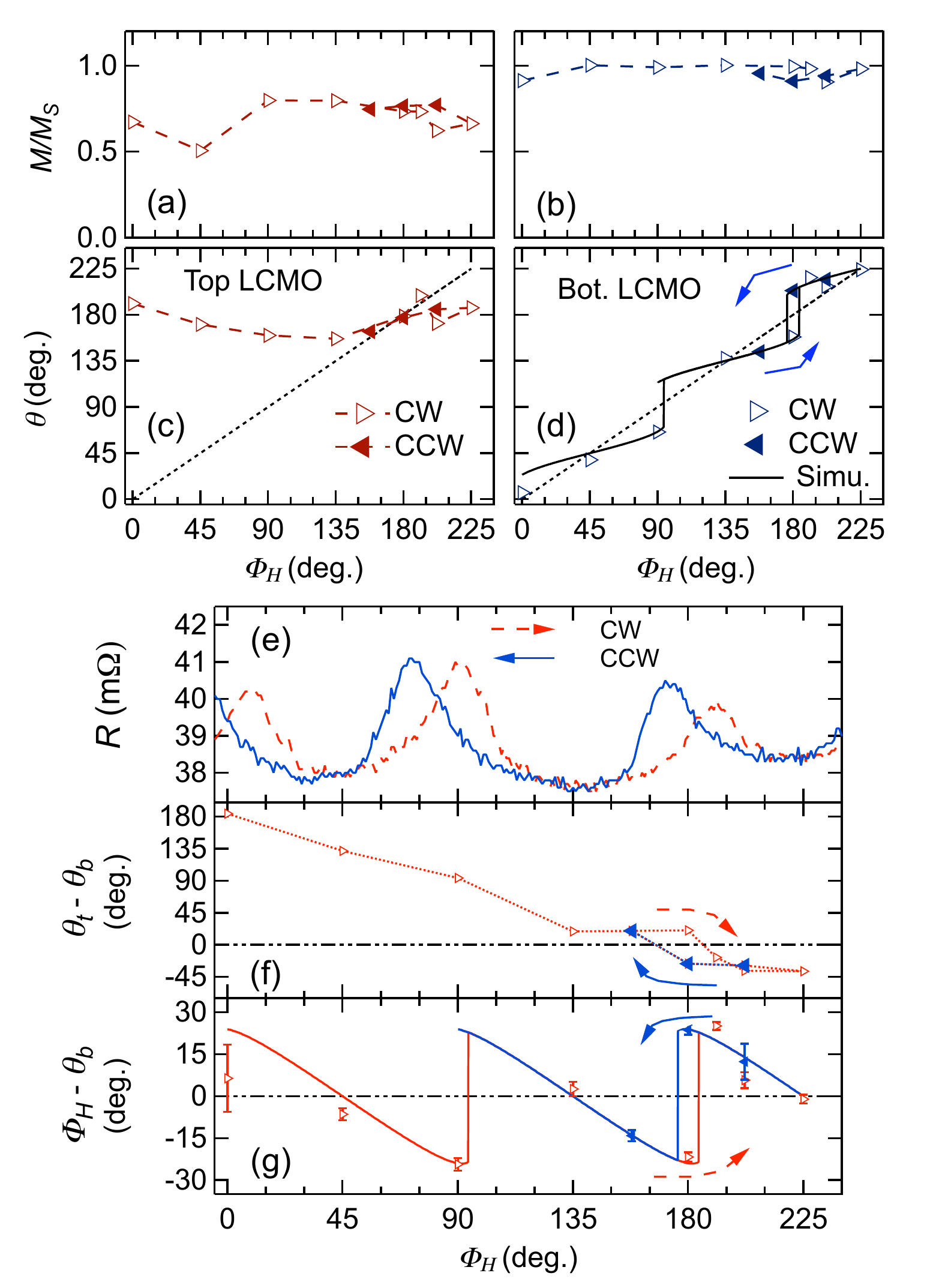}
	\caption{\label{Fig:Theta}~(Color online) Normalized amplitudes ($M/M_{S}$) and directions ($\theta_t$ and $\theta_b$) for the top (a, c) and bottom (b, d) LCMO magnetizations  during rotation, as determined from our PNR experiments. The dotted lines in (c) and (d) show the field direction $\Phi_{H}$. (e) Angular dependence of MR (same as Fig.~\ref{Fig:MR}(b)). Relative orientations (f) between $\vec{M}_{t}$ and $\vec{M}_{b}$, and (g) between $\vec{M}_{b}$ and $\vec{H}_a$,  determined from the PNR (triangles), respectively. The solid lines in (d) and (g) show the calculated results based on the energy minimization. Clearly, $\vec{M}_{b}$  is parallel to $\vec{H}_{a}$ when the field is along an easy axis .} 
\end{figure}

Next we study the response of the top and bottom layer magnetization during rotation  of the magnetic field.  A $150$~Oe field was applied along $0^\circ$ after having saturated the sample along $180^\circ$. The experiments were then conducted at the following field directions sequentially: $0^\circ$, $45^\circ$, $90^\circ$, $135^\circ$, $180^\circ$, $191^\circ$, $202^\circ$, $225^\circ$; and then $202^\circ$, $180^\circ$ and $158^\circ$. In contrast to the case for saturation, there the spin-flip reflectivities ($R^{SF}$) are  non-zero. $R^{SF}$ is sensitive to the square of the components of the magnetization perpendicular to the field direction~\cite{FelcherRSI1987, MajkrzakPhysB1996, Fitz2005}. Figures~\ref{Fig:PNR}(b)-(d) show some representative data. $R^{SF}$ is high at $90^\circ$; it becomes lower at $135^\circ$ and slightly increases again at $180^\circ$. We determine the direction and magnitude of the magnetizations for the top ($\vec{M}_{t}$) and bottom ($\vec{M}_{b}$) LCMO layers independently at each field direction by fitting $R^{++}$, $R^{--}$ and $R^{SF}$ all together. Figures~\ref{Fig:Theta}(a)-(d) shows the $\vec{M}_{t}$ and $\vec{M}_{b}$ obtained from the best fit as a function of the field direction. The amplitudes are normalized to their respective saturation magnetizations. $ \theta_{t} $ and $ \theta_{b} $ are the directions of  $\vec{M}_{t}$ and $\vec{M}_{b}$, respectively,  with respect to the [100] axis. The magnetic field affects the magnitude of the top layer magnetization, but not its direction.  This implies the top layer breaks up into domains. On the other hand, the magnetic field affects the orientation of the bottom layer magnetization but not its magnitude.  Thus the bottom layer apparently rotates in response to field.

Because $\vec{M}_{b}$ keeps the saturation amplitude during the rotation, we use the coherent rotation model to estimate its expected direction to achieve a more detailed picture of its magnetization structure during rotation. We consider the Zeeman energy and the anisotropy energy in the free energy, i.e.  $F=-~\vec{M}_{b} \cdot \vec{H}_{a} + K_{4} \times  \cos^{2}( 2 \theta_{b} ) $, where $M = M_{S} = 540$~emu/cm$^3$,  $H$ = 150~Oe, and $K_{4}$ is the biaxial magnetocrystalline anisotropy~\cite{VisaniPRB2011}. $\theta_{b}$ is computed via minimizing the free energy. As shown in Fig.~\ref{Fig:Theta}(d), the calculated values well match the PNR results with $K_{4} =1.6 \times 10^4$~erg/cm$^3$.  The only exception is at $0^\circ$ because of its different magnetic history (field sweeping rather than rotation).  Clearly, $\vec{M}_{b}$ is parallel to $\vec{H}_{a}$ when $\vec{H}_{a}$ is along an easy axis. At the same time, the angle between $\vec{M_{b}}$ and $\vec{H}_{a}$ reaches a local maximum when the field slightly passes a hard axis, and it shows a hysteresis between clockwise and counter-clockwise rotations. 

Figure~\ref{Fig:Theta}(f) shows the relative orientation between $\vec{M}_{t}$ and $\vec{M}_{b}$ as the field was rotated. $\vec{M}_{t}$ and $\vec{M}_{b}$ are nearly antiparallel (AP) when $\Phi_{H}=0^\circ$, and parallel (P) when $\Phi_{H}=135^\circ$. More importantly, the difference between the orientations of $M_t$ and $M_b$ decreases monotonically as $\Phi_{H}$ increases from $0^\circ$ to $135^\circ$. Despite of a sign change, the amplitude of the relative orientation changes little between $\Phi_{H} = 135^\circ$ and $225^\circ$. If the magnetization alignment governed the MR monotonically, such as for the spin-dependent scattering, then the MR would show no oscillations between $0^\circ$~and $135^\circ$ and change little between $135^\circ$ and $225^\circ$. These are obviously in contrast to the MR data shown in Fig.~\ref{Fig:Theta}(e). Therefore, our results exclude many scenarios that are based on the concept of the mutual magnetization alignment.

\begin{figure}[tb]
	\centering
		\includegraphics[width=0.42\textwidth]{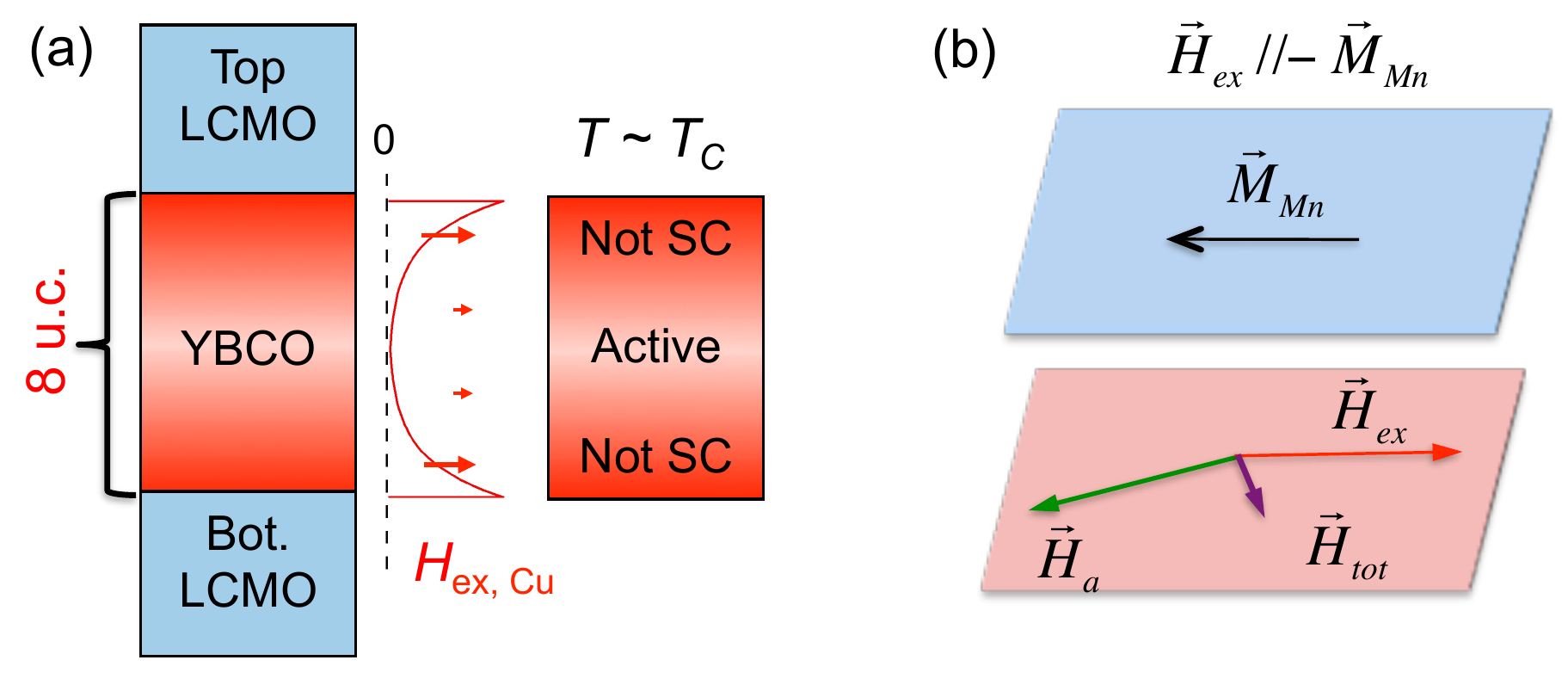}
	\caption{\label{Fig:Htot}~(Color online) (a) A schematic picture of the induced exchange field $\vec{H}_{ex}$ in YBCO.  $H_{ex}$ decays exponentially from the interface.   When $T \sim T_{C}$,  only the central YBCO undergoes the superconducting transition and therefore dominates the transport properites, because the superconductivity in the interfacial YBCO is strongly suppressed. (b) $\vec{H}_{ex}$ is antiparallel to the magnetization of the adjacent LCMO layer ($\vec{M}_{Mn}$). Meanwhile, the superconductivity in central YBCO is subject to $\vec{H}_{tot}$ that results from the superposition of $\vec{H}_{a}$ and (the tail of) $\vec{H}_{ex}$.  Therefore, the relative alignment between $\vec{H}_{a}$ and $\vec{M}_{Mn}$ plays a key role in controlling the superconductivity. } 
\end{figure}

On the other hand, the interplay between $\vec{H}_{ex}$ (from the bottom interface) and $\vec{H}_{a}$ is able to explain the oscillations of the MR with $\Phi_H$. $\vec{H}_{ex}$ is on the order of a few of hundreds of Tesla in the first interfacial YBCO unit cell~\cite{ChakhalianNatPhy2006}.  Salafranca and Okamoto have shown that $\vec{H}_{ex}$ decays exponentially from the interface with an attenuation length less than 1 u.c. and does not quite reach the center of 8~u.c. thick YBCO when $T = T_C$; therefore,  $\vec{H}_{ex}$'s from the top and bottom interfaces influence the superconductivity independently~\cite{SalafrancaPRL2010}. At the same time, both the coherence length and the mean free path are $\leq$~1~u.c. along the $c$-axis in YBCO ~\cite{WorthingtonPRL1987, LiPRL1990}. Therefore, we view the 8~u.c.~YBCO layer as a few of parallel sublayers for simplicity. This situation is shown in Fig.~\ref{Fig:Htot}(a). $\vec{H}_{ex}$ in the central YBCO is much weaker than in the interfacial one, so that only the central YBCO becomes superconduting and dominates the resistance of the trilayers when  $T = T_C$. The superconductivity in central YBCO is subject to $\vec{H}_{tot}$ that results from the superposition of $\vec{H}_{a}$ and (the tail of) $\vec{H}_{ex}$. The relative alignment between $\vec{H}_{a}$ and $\vec{H}_{ex}$ determines the amplitude of $\vec{H}_{tot}$ during the field rotation (see Fig.~\ref{Fig:Htot}(b)). The change of $H_{tot}$ is on the same order of the applied field (150 Oe) during the rotation, slightly shifting the superconducting transition curves and giving rise to a small but observable MR.  When $\vec{H}_{ex}$ and $\vec{H}_{a}$ are antiparallel, they effectively cancel each other. Since $\vec{H}_{ex}$ is antiparallel to $\vec{M}_{Mn}$, $H_{tot}$ is weakest when $\vec{H}_{a}$ is parallel to $\vec{M}_{Mn}$, which gives rise to a low resistance state. As shown in Fig.~\ref{Fig:Theta}(g),  $\vec{H}_{a}$ is parallel to $\vec{M}_{b}$   when $\vec{H}_{a}$ is along an easy axis direction with corresponding resistance minima. At intermediate angles, ${H}_{tot}$ varies and so does MR. This explains the four-fold symmetry of MR. At the same time, the angular hysteresis of $\vec{M}_{b}$ with respect to the field direction gives rise to the hysteresis in both ${H}_{tot}$ and MR. 

From Salafranca-Okamoto's theory, we also expect an unidirectional offset in MR due to the balance between the external field and the exchange field from the top surface since $\vec{M}_{t}$ retains the initial saturation direction. The sample used in this PNR study does not show this expected offset and the reason is unclear. One possibility is that as a result of the top LCMO layer breaking down into domains, the effect is compromised. However, such an offset is observed in other samples. Figure~1(b) in Ref.~\cite{NemesAPL2010} is an example. It clearly shows that, beside the hysteretic four-fold symmetry, there is a unidirectional offset in MR along the initial saturation direction.

A final remark concerns the effect of stray fields created by domain walls, of ferromagnetically coupled face-to-face domains in the two FM layers. It has been argued that the magnetic flux closure of the enhanced stray field at domain walls through the SC will cause a large MR~\cite{ZhuPRL2009, StamopoulosSST2007}. This does not occur in our rotation experiment at 150 Oe because the bottom LCMO maintains its saturation magnetization. However, we do observe additional MR features due to the effect of stray fields in other rotation sequences~\cite{SM}. 

In summary, we have shown that the interfacial electronic reconstruction controls the inverse spin switch behaviour of half metal-superconductor oxide spin valves. The angular dependence of MR in LYL trilayers along the superconducting transition displays symmetry features that are not correlated with the relative alignment between the two FM's magnetizations, which rules out many MR scenarios proposed so far.  Rather the field dependence of the MR is explained by the interplay between the applied field and (the tail of) the induced exchange field on YBCO coming from the electronic reconstruction at the LCMO/YBCO interface. Since the inverse spin switch in LYL is now demonstrated to be governed by interfacial electronic reconstruction and not shape dependent micromagnetic effects, we expect it to survive miniaturization to the nanoscale. 

\begin{acknowledgments}
We thank S. Okamoto and J. Salafranca for valuable discussions. Research at Argonne National Laboratory was supported by the U.S. Department of Energy, Office of Basic Energy Sciences, Division of Materials Sciences and Engineering under Award No.DE-AC02-06CH11357. Work at UCM was supported by Spanish MICINN Grant MAT 2011 27470, Consolider Ingenio CSD2009-00013 (IMAGINE), CAM S2009-MAT 1756 (PHAMA). This work has benefited from the use of the Lujan Neutron Scattering Center at LANSCE, which is funded by the Department of Energy's Office of Basic Energy Sciences. Los Alamos National Laboratory is operated by Los Alamos National Security LLC under DOE through Contract No. DE-AC52-06NA25396. 
\end{acknowledgments}

\section {\large{Supplemental Material}}
\section{Complementary studies of x-ray and polarized neturon reflectometry}

\renewcommand{\thefigure}{\arabic{figure}SM}

\begin{figure}[b]
	\begin{centering}
	\includegraphics[width= 0.38 \textwidth]{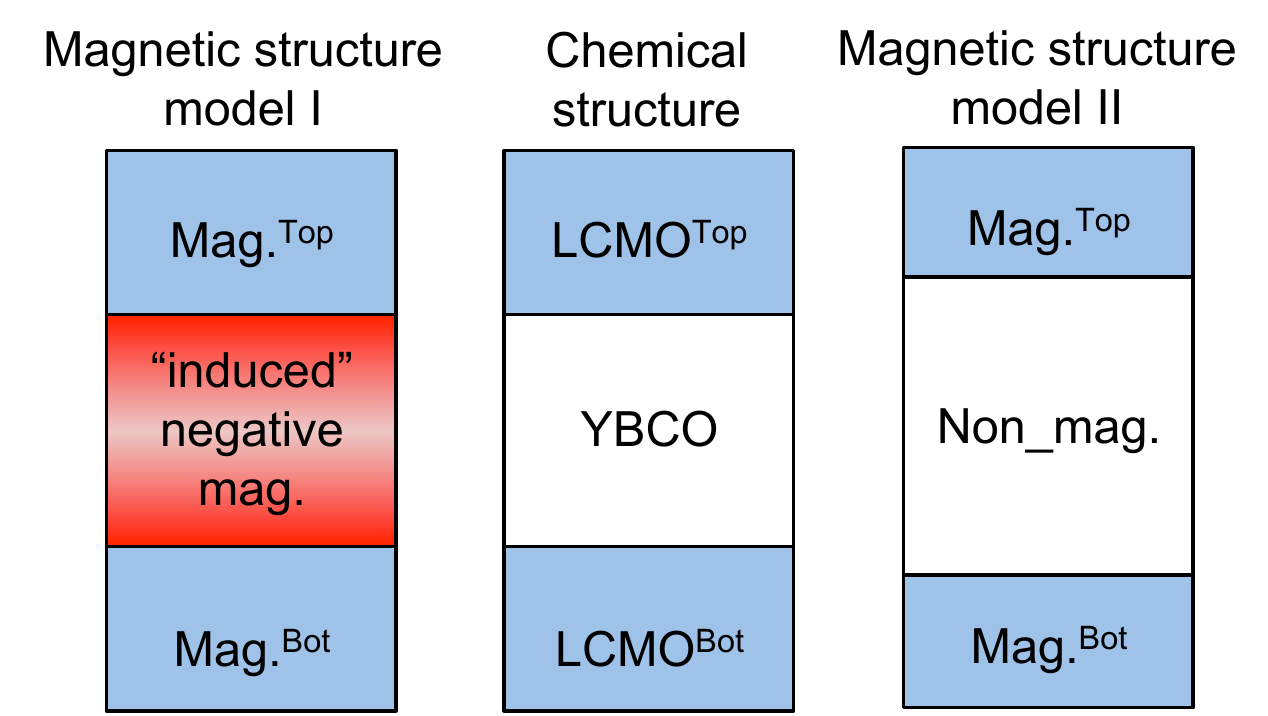}
	  \caption{\label{Fig:Model} The middle panel illustrates the chemical structure model, and the left and right panels show the two magnetic structure models used to describe the LCMO/YBCO interfaces. Model I (left panel) includes a non-zero magnetization in YBCO. We assume the magnetization in YBCO exponentially decays from the interfaces, as illustrated by the color scale.  Model II (right panel) considers  the magnetization suppression  in LCMO at the interfaces.}
	\end{centering} 
\end{figure}

We have performed complementary studies of x-ray and polarized neturon reflectometry to determine the sample's chemical and magnetic structures. The x-ray reflectivity (XRR) data were collected at room temperature and the polarized neturon reflectivity (PNR) data in saturation were collected at 26 K in a 5 kOe field. We simulate XRR and PNR based on the Parratt formalism~\cite{ParrattPR1954} and model a rough interface as a sequence of very thin slices whose scattering length densities (SLDs) vary, following an error function so as to interpolate between adjacent layers. The neutron magnetic SLD is directly proportional to magnetization through the coefficient $2.853 \times 10^{-9}$~\AA$^{-2}$cm$^{3}$/emu. We have considered two models to describe the magnetic structure at the La$_{0.7}$Ca$_{0.3}$MnO$_3$ (LCMO)/YBa$_2$Cu$_3$O$_{7-\delta}$ (YBCO) interfaces, as illustrated in Fig.~\ref{Fig:Model}. 

 \begin{figure}
	\begin{centering}
	\includegraphics[width=0.5 \textwidth]{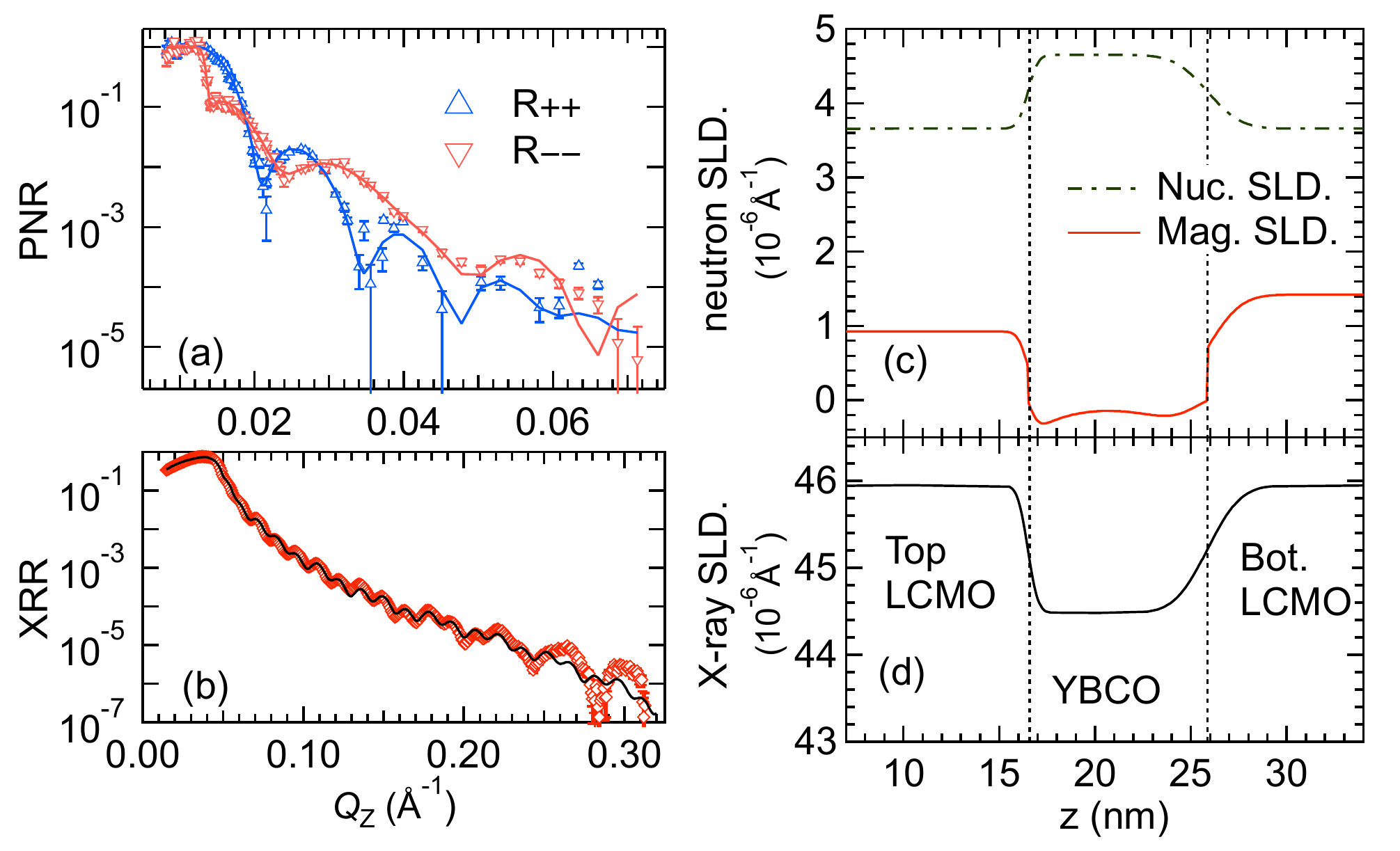}
	  \caption{\label{Fig:InvM}  Results from model I. (a), (b) Data and the best fit for PNR and XRR, respectively. (c), (d) Depth profiles of the neutron nuclear and magnetic SLDs, and the real part of the x-ray SLD, respectively.}
	\end{centering}
\end{figure}

 \begin{figure}
	\begin{centering}
	\includegraphics[width=0.5 \textwidth]{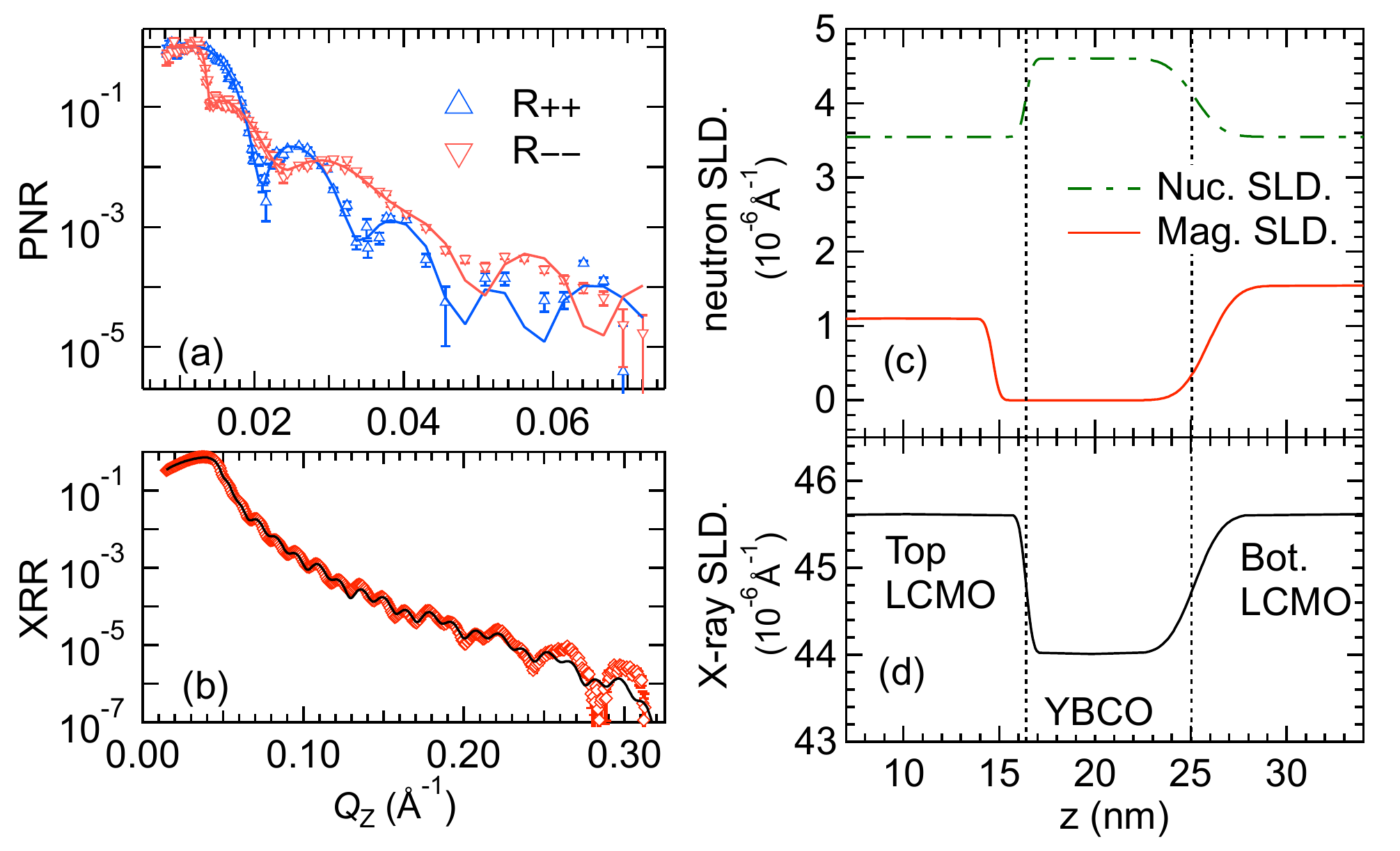}
	  \caption{\label{Fig:DeadM}  Results from model II. (a), (b) Data and the best fit for PNR and XRR, respectively. (c), (d) Depth profiles of the neutron nuclear and magnetic SLDs, and the real part of the x-ray SLD, respectively. }	  
	\end{centering}
\end{figure}

Model I considers the ``induced" negative magnetization in YBCO~\cite{ChakhalianNatPhy2006, SalafrancaPRL2010, WernerPRB2010} (the left panel in Fig.~\ref{Fig:Model}). We assume that the amplitude of the induced magnetization exponentially decays from the LCMO/YBCO  interfaces~\cite{StahnPRB2005, SalafrancaPRL2010}. The results are shown in Fig.~\ref{Fig:InvM}. After taking into account the interface roughness effect, the largest magnetization amplitude in YBCO is $\sim -10\%$ of that in LCMOs and the decay length of the induced magnetization is 1.7~nm. 

Model II considers the suppression of magnetization inside LCMO at both LCMO/YBCO interfaces~\cite{HoffmanPRB2005, LuoPRL2008} (the right panel in Fig.~\ref{Fig:Model}).  In order to reduce the fitting parameters, we assume that the LCMO's magnetization near the interface is either zero or the same as the central LCMO's magnetization.   The results are shown in Fig.~\ref{Fig:DeadM}. The best fit shows that the magnetization suppression regions are of 1.8 and 0.6~nm at the top and bottom interfaces, respectively. 
 
Similar to previous PNR studies on LCMO/YBCO heterostructrues in a limited $Q_z$ range~\cite{StahnPRB2005}, Model I yields a better fit for $Q_z$ between 0.045 and 0.055~\AA$^{-1}$ and model II yields a better fit for $Q_z$ between 0.035 to 0.045~\AA$^{-1}$~\cite{bkg}.  Therefore, our PNR experiments are not sufficient to resolve the subtleness of the magnetization profile at the interfaces. However, both models show that the top LCMO has a lower saturation magnetization than the bottom one does. We have used the model II to analyze the low field PNR data due to its relative simplicity.  It is worth noting that the result from model II only suggests a magnetization suppression, rather than magnetic dead layers in LCMOs near the interfaces. The saturation magnetizations from the best fit using the model II are $\sim380$ and  $\sim 540$~emu/cm$^3$ for the top and bottom LCMOs, respectively.

\section{Additional transport experiments}

\subsection{Angular dependence of magnetoresistance at 30 Oe}
\begin{figure}[tb]
	\centering
		\includegraphics[width=0.3\textwidth]{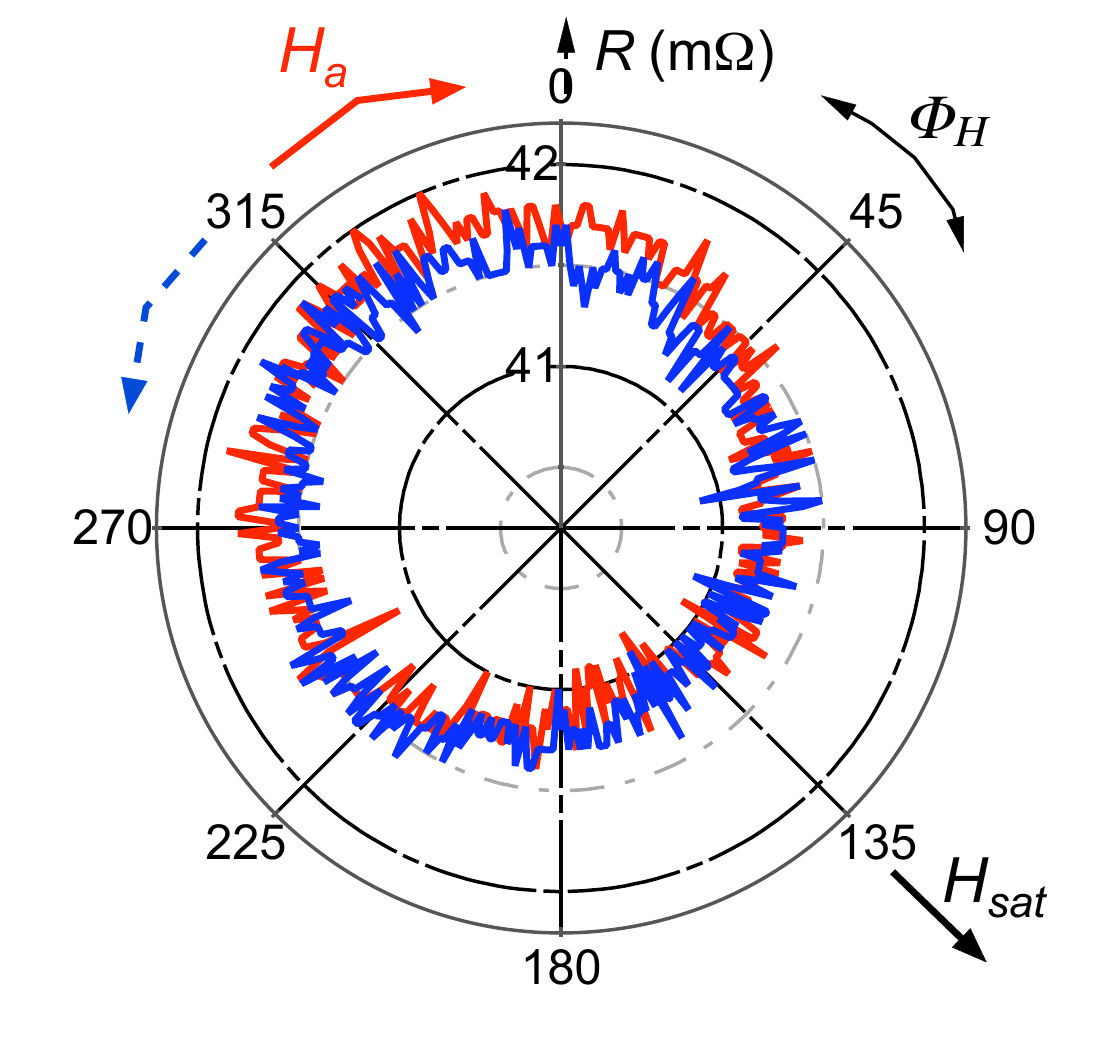}
	\caption{\label{Fig:MR30Oe} Angular dependence of the resistance in a 30 Oe in-plane field. Data were taken at 26 K.} 
\end{figure}

From Salafranca-Okamoto's theory, if the applied field is so low that both $\vec{M}_{t}$ and $\vec{M}_{b}$ retain their directions during rotation, we expect a resistance minimum when the field is along the initial saturation direction, and a resistance maximum antiparallel to it. This is consistent with the experimental results in Fig.~\ref{Fig:MR30Oe}. The data shows the angular dependence of the resistance at 30 Oe, which is much lower than coercivities of both LCMOs. The experimental sequence is similar to that mentioned in paper, but  a 30~Oe rotation field is used after having saturated the sample along 135$^\circ$. 

\subsection{Stray field effects}

\begin{figure}[tb]
	\centering
		\includegraphics[width=0.32\textwidth]{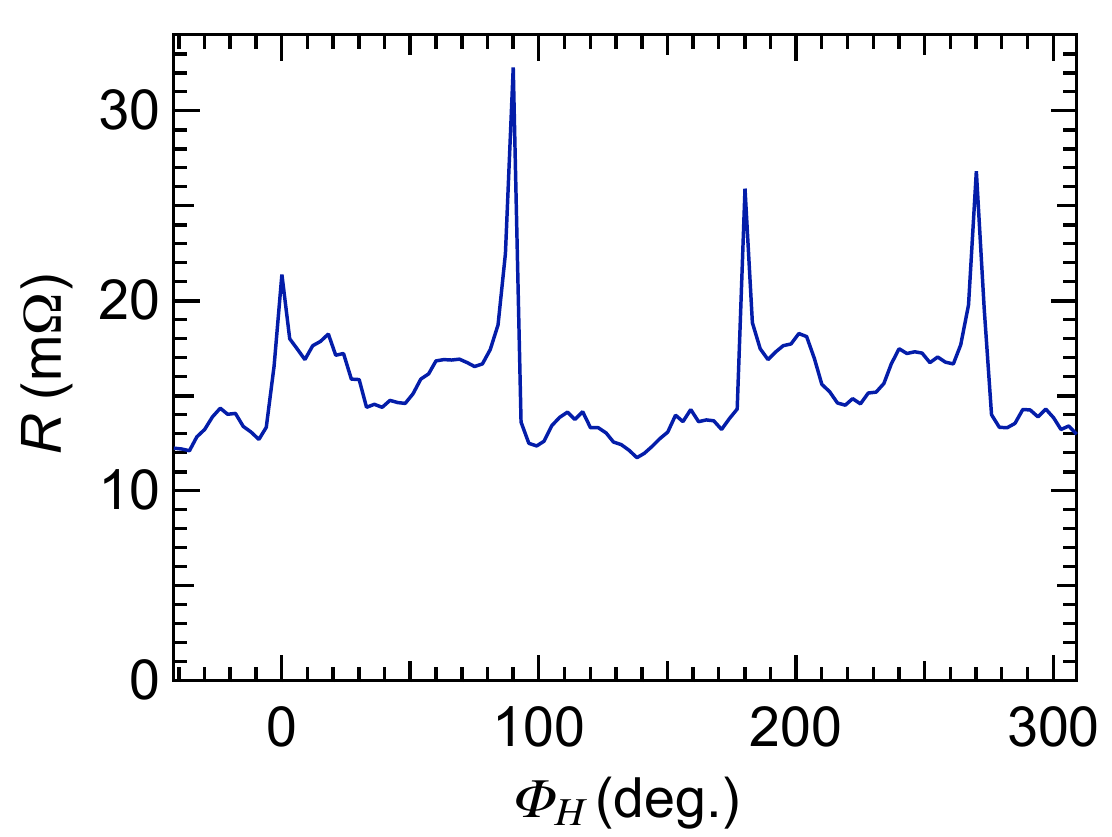}
	\caption{\label{Fig:RH} Resistances at 150 Oe extracted individually from several ascending scans from field-sweeps along various orientations. The MR has peaks along the hard axis directions. The data were collected with a different sample. } 
\end{figure}

Other field rotation sequences are specifically designed to probe the effect of stray fields due to ferromagnetically coupled face-to-face domains in the two LCMO layers, since these enhance the normal component of the stray field at the correlated domain-walls in the two FM layers~\cite{ZhuPRL2009}. At fixed angles we sweep the field from negative to positive saturation  (10 kOe) and back. Figure~\ref{Fig:RH} shows resistance versus field direction at positive 150 Oe after negative saturation as a function of the in-plane field direction. In addition to the hysteretic four-fold  MR curves that appear in the fixed field rotation, there are four enhanced  MR peaks when the field is swept along hard axis directions, which is the signature of the effect from the stray fields.  When the field is swept along neither  a hard axis nor an easy axis, there  is a preferred rotation  direction; therefore, the multidomain formation  is suppressed,  so is the stray  field effect. When the field is swept along a hard axis, magnetization reversal occurs by domain formation and these rotate gradually in opposite directions away from the hard axis. On the other hand, when the field is swept along an easy axis, FM domains are aligned in opposite directions along the easy axis. However, the multidomain state exists in a much larger field range along the hard axis than it does along the easy axis. Because the top  and bottom  LCMOs have very different coercitvities, it is much more favorable to form coupled face-to-face domains  in the  two LCMOs in a hard  axis field sweep than in an easy axis field sweep.  This suggests that the four MR peaks along the hard axes are due to the effect of stray fields.   

Nevertheless, there  are no sharp  MR peaks during  the  field rotation experiments. Our PNR experiments also indicate that  the  bottom LCMO has a single-domain like magnetization during rotation. Therefore,  we conclude that  the effect of stray  field is suppressed  during the field rotation  at 150 Oe.

\bibliography{LYL_Rot}	

\begin{thebibliography}{32}%
\makeatletter
\providecommand \@ifxundefined [1]{%
 \@ifx{#1\undefined}
}%
\providecommand \@ifnum [1]{%
 \ifnum #1\expandafter \@firstoftwo
 \else \expandafter \@secondoftwo
 \fi
}%
\providecommand \@ifx [1]{%
 \ifx #1\expandafter \@firstoftwo
 \else \expandafter \@secondoftwo
 \fi
}%
\providecommand \natexlab [1]{#1}%
\providecommand \enquote  [1]{``#1''}%
\providecommand \bibnamefont  [1]{#1}%
\providecommand \bibfnamefont [1]{#1}%
\providecommand \citenamefont [1]{#1}%
\providecommand \href@noop [0]{\@secondoftwo}%
\providecommand \href [0]{\begingroup \@sanitize@url \@href}%
\providecommand \@href[1]{\@@startlink{#1}\@@href}%
\providecommand \@@href[1]{\endgroup#1\@@endlink}%
\providecommand \@sanitize@url [0]{\catcode `\\12\catcode `\$12\catcode
  `\&12\catcode `\#12\catcode `\^12\catcode `\_12\catcode `\%12\relax}%
\providecommand \@@startlink[1]{}%
\providecommand \@@endlink[0]{}%
\providecommand \url  [0]{\begingroup\@sanitize@url \@url }%
\providecommand \@url [1]{\endgroup\@href {#1}{\urlprefix }}%
\providecommand \urlprefix  [0]{URL }%
\providecommand \Eprint [0]{\href }%
\providecommand \doibase [0]{http://dx.doi.org/}%
\providecommand \selectlanguage [0]{\@gobble}%
\providecommand \bibinfo  [0]{\@secondoftwo}%
\providecommand \bibfield  [0]{\@secondoftwo}%
\providecommand \translation [1]{[#1]}%
\providecommand \BibitemOpen [0]{}%
\providecommand \bibitemStop [0]{}%
\providecommand \bibitemNoStop [0]{.\EOS\space}%
\providecommand \EOS [0]{\spacefactor3000\relax}%
\providecommand \BibitemShut  [1]{\csname bibitem#1\endcsname}%
\let\auto@bib@innerbib\@empty
\bibitem [{\citenamefont {Ohtomo}\ and\ \citenamefont
  {Hwang}(2004)}]{OhtomoNature2004}%
  \BibitemOpen
  \bibfield  {author} {\bibinfo {author} {\bibfnamefont {A.}~\bibnamefont
  {Ohtomo}}\ and\ \bibinfo {author} {\bibfnamefont {H.}~\bibnamefont {Hwang}},\
  }\href {\doibase 10.1038/nature02308} {\bibfield  {journal} {\bibinfo
  {journal} {NATURE}\ }\textbf {\bibinfo {volume} {427}},\ \bibinfo {pages}
  {423} (\bibinfo {year} {2004})}\BibitemShut {NoStop}%
\bibitem [{\citenamefont {Reyren}\ \emph {et~al.}(2007)\citenamefont {Reyren},
  \citenamefont {Thiel}, \citenamefont {Caviglia}, \citenamefont {Kourkoutis},
  \citenamefont {Hammerl}, \citenamefont {Richter}, \citenamefont {Schneider},
  \citenamefont {Kopp}, \citenamefont {Ruetschi}, \citenamefont {Jaccard},
  \citenamefont {Gabay}, \citenamefont {Muller}, \citenamefont {Triscone},\
  and\ \citenamefont {Mannhart}}]{ReyrenScience2007}%
  \BibitemOpen
  \bibfield  {author} {\bibinfo {author} {\bibfnamefont {N.}~\bibnamefont
  {Reyren}}, \bibinfo {author} {\bibfnamefont {S.}~\bibnamefont {Thiel}},
  \bibinfo {author} {\bibfnamefont {A.~D.}\ \bibnamefont {Caviglia}}, \bibinfo
  {author} {\bibfnamefont {L.~F.}\ \bibnamefont {Kourkoutis}}, \bibinfo
  {author} {\bibfnamefont {G.}~\bibnamefont {Hammerl}}, \bibinfo {author}
  {\bibfnamefont {C.}~\bibnamefont {Richter}}, \bibinfo {author} {\bibfnamefont
  {C.~W.}\ \bibnamefont {Schneider}}, \bibinfo {author} {\bibfnamefont
  {T.}~\bibnamefont {Kopp}}, \bibinfo {author} {\bibfnamefont {A.-S.}\
  \bibnamefont {Ruetschi}}, \bibinfo {author} {\bibfnamefont {D.}~\bibnamefont
  {Jaccard}}, \bibinfo {author} {\bibfnamefont {M.}~\bibnamefont {Gabay}},
  \bibinfo {author} {\bibfnamefont {D.~A.}\ \bibnamefont {Muller}}, \bibinfo
  {author} {\bibfnamefont {J.-M.}\ \bibnamefont {Triscone}}, \ and\ \bibinfo
  {author} {\bibfnamefont {J.}~\bibnamefont {Mannhart}},\ }\href {\doibase
  10.1126/science.1146006} {\bibfield  {journal} {\bibinfo  {journal}
  {SCIENCE}\ }\textbf {\bibinfo {volume} {317}},\ \bibinfo {pages} {1196}
  (\bibinfo {year} {2007})}\BibitemShut {NoStop}%
\bibitem [{\citenamefont {Brinkman}\ \emph {et~al.}(2007)\citenamefont
  {Brinkman}, \citenamefont {Huijben}, \citenamefont {Van~Zalk}, \citenamefont
  {Huijben}, \citenamefont {Zeitler}, \citenamefont {Maan}, \citenamefont
  {Van~der Wiel}, \citenamefont {Rijnders}, \citenamefont {Blank},\ and\
  \citenamefont {Hilgenkamp}}]{BrinkmanNatMat2007}%
  \BibitemOpen
  \bibfield  {author} {\bibinfo {author} {\bibfnamefont {A.}~\bibnamefont
  {Brinkman}}, \bibinfo {author} {\bibfnamefont {M.}~\bibnamefont {Huijben}},
  \bibinfo {author} {\bibfnamefont {M.}~\bibnamefont {Van~Zalk}}, \bibinfo
  {author} {\bibfnamefont {J.}~\bibnamefont {Huijben}}, \bibinfo {author}
  {\bibfnamefont {U.}~\bibnamefont {Zeitler}}, \bibinfo {author} {\bibfnamefont
  {J.~C.}\ \bibnamefont {Maan}}, \bibinfo {author} {\bibfnamefont {W.~G.}\
  \bibnamefont {Van~der Wiel}}, \bibinfo {author} {\bibfnamefont
  {G.}~\bibnamefont {Rijnders}}, \bibinfo {author} {\bibfnamefont {D.~H.~A.}\
  \bibnamefont {Blank}}, \ and\ \bibinfo {author} {\bibfnamefont
  {H.}~\bibnamefont {Hilgenkamp}},\ }\href {\doibase 10.1038/nmat1931}
  {\bibfield  {journal} {\bibinfo  {journal} {NATURE MATERIALS}\ }\textbf
  {\bibinfo {volume} {6}},\ \bibinfo {pages} {493} (\bibinfo {year}
  {2007})}\BibitemShut {NoStop}%
\bibitem [{\citenamefont {Salafranca}\ and\ \citenamefont
  {Okamoto}(2010)}]{SalafrancaPRL2010}%
  \BibitemOpen
  \bibfield  {author} {\bibinfo {author} {\bibfnamefont {J.}~\bibnamefont
  {Salafranca}}\ and\ \bibinfo {author} {\bibfnamefont {S.}~\bibnamefont
  {Okamoto}},\ }\href {\doibase 10.1103/PhysRevLett.105.256804} {\bibfield
  {journal} {\bibinfo  {journal} {Phys. Rev. Lett.}\ }\textbf {\bibinfo
  {volume} {105}},\ \bibinfo {pages} {256804} (\bibinfo {year}
  {2010})}\BibitemShut {NoStop}%
\bibitem [{\citenamefont {Chakhalian}\ \emph {et~al.}(2006)\citenamefont
  {Chakhalian}, \citenamefont {Freeland}, \citenamefont {Srajer}, \citenamefont
  {Strempfer}, \citenamefont {Khaliullin}, \citenamefont {Cezar}, \citenamefont
  {Charlton}, \citenamefont {Dalgliesh}, \citenamefont {Bernhard},
  \citenamefont {Cristiani} \emph {et~al.}}]{ChakhalianNatPhy2006}%
  \BibitemOpen
  \bibfield  {author} {\bibinfo {author} {\bibfnamefont {J.}~\bibnamefont
  {Chakhalian}}, \bibinfo {author} {\bibfnamefont {J.}~\bibnamefont
  {Freeland}}, \bibinfo {author} {\bibfnamefont {G.}~\bibnamefont {Srajer}},
  \bibinfo {author} {\bibfnamefont {J.}~\bibnamefont {Strempfer}}, \bibinfo
  {author} {\bibfnamefont {G.}~\bibnamefont {Khaliullin}}, \bibinfo {author}
  {\bibfnamefont {J.}~\bibnamefont {Cezar}}, \bibinfo {author} {\bibfnamefont
  {T.}~\bibnamefont {Charlton}}, \bibinfo {author} {\bibfnamefont
  {R.}~\bibnamefont {Dalgliesh}}, \bibinfo {author} {\bibfnamefont
  {C.}~\bibnamefont {Bernhard}}, \bibinfo {author} {\bibfnamefont
  {G.}~\bibnamefont {Cristiani}},  \emph {et~al.},\ }\href {\doibase
  10.1038/nphys272} {\bibfield  {journal} {\bibinfo  {journal} {Nature Phys.}\
  }\textbf {\bibinfo {volume} {2}},\ \bibinfo {pages} {244} (\bibinfo {year}
  {2006})}\BibitemShut {NoStop}%
\bibitem [{\citenamefont {Werner}\ \emph {et~al.}(2010)\citenamefont {Werner},
  \citenamefont {Raisch}, \citenamefont {Ruosi}, \citenamefont {Davidson},
  \citenamefont {Nagel}, \citenamefont {Merz}, \citenamefont {Schuppler},
  \citenamefont {Glaser}, \citenamefont {Fujii}, \citenamefont {Chass\'e},
  \citenamefont {Kleiner},\ and\ \citenamefont {Koelle}}]{WernerPRB2010}%
  \BibitemOpen
  \bibfield  {author} {\bibinfo {author} {\bibfnamefont {R.}~\bibnamefont
  {Werner}}, \bibinfo {author} {\bibfnamefont {C.}~\bibnamefont {Raisch}},
  \bibinfo {author} {\bibfnamefont {A.}~\bibnamefont {Ruosi}}, \bibinfo
  {author} {\bibfnamefont {B.~A.}\ \bibnamefont {Davidson}}, \bibinfo {author}
  {\bibfnamefont {P.}~\bibnamefont {Nagel}}, \bibinfo {author} {\bibfnamefont
  {M.}~\bibnamefont {Merz}}, \bibinfo {author} {\bibfnamefont {S.}~\bibnamefont
  {Schuppler}}, \bibinfo {author} {\bibfnamefont {M.}~\bibnamefont {Glaser}},
  \bibinfo {author} {\bibfnamefont {J.}~\bibnamefont {Fujii}}, \bibinfo
  {author} {\bibfnamefont {T.}~\bibnamefont {Chass\'e}}, \bibinfo {author}
  {\bibfnamefont {R.}~\bibnamefont {Kleiner}}, \ and\ \bibinfo {author}
  {\bibfnamefont {D.}~\bibnamefont {Koelle}},\ }\href {\doibase
  10.1103/PhysRevB.82.224509} {\bibfield  {journal} {\bibinfo  {journal} {Phys.
  Rev. B}\ }\textbf {\bibinfo {volume} {82}},\ \bibinfo {pages} {224509}
  (\bibinfo {year} {2010})}\BibitemShut {NoStop}%
\bibitem [{\citenamefont {Visani}\ \emph {et~al.}(2011)\citenamefont {Visani},
  \citenamefont {Tornos}, \citenamefont {Nemes}, \citenamefont {Rocci},
  \citenamefont {Leon}, \citenamefont {Santamaria}, \citenamefont
  {te~Velthuis}, \citenamefont {Liu}, \citenamefont {Hoffmann}, \citenamefont
  {Freeland}, \citenamefont {Garcia-Hernandez}, \citenamefont {Fitzsimmons},
  \citenamefont {Kirby}, \citenamefont {Varela},\ and\ \citenamefont
  {Pennycook}}]{VisaniPRB2011}%
  \BibitemOpen
  \bibfield  {author} {\bibinfo {author} {\bibfnamefont {C.}~\bibnamefont
  {Visani}}, \bibinfo {author} {\bibfnamefont {J.}~\bibnamefont {Tornos}},
  \bibinfo {author} {\bibfnamefont {N.~M.}\ \bibnamefont {Nemes}}, \bibinfo
  {author} {\bibfnamefont {M.}~\bibnamefont {Rocci}}, \bibinfo {author}
  {\bibfnamefont {C.}~\bibnamefont {Leon}}, \bibinfo {author} {\bibfnamefont
  {J.}~\bibnamefont {Santamaria}}, \bibinfo {author} {\bibfnamefont {S.~G.~E.}\
  \bibnamefont {te~Velthuis}}, \bibinfo {author} {\bibfnamefont
  {Y.}~\bibnamefont {Liu}}, \bibinfo {author} {\bibfnamefont {A.}~\bibnamefont
  {Hoffmann}}, \bibinfo {author} {\bibfnamefont {J.~W.}\ \bibnamefont
  {Freeland}}, \bibinfo {author} {\bibfnamefont {M.}~\bibnamefont
  {Garcia-Hernandez}}, \bibinfo {author} {\bibfnamefont {M.~R.}\ \bibnamefont
  {Fitzsimmons}}, \bibinfo {author} {\bibfnamefont {B.~J.}\ \bibnamefont
  {Kirby}}, \bibinfo {author} {\bibfnamefont {M.}~\bibnamefont {Varela}}, \
  and\ \bibinfo {author} {\bibfnamefont {S.~J.}\ \bibnamefont {Pennycook}},\
  }\href {\doibase 10.1103/PhysRevB.84.060405} {\bibfield  {journal} {\bibinfo
  {journal} {Phys. Rev. B}\ }\textbf {\bibinfo {volume} {84}},\ \bibinfo
  {pages} {060405} (\bibinfo {year} {2011})}\BibitemShut {NoStop}%
\bibitem [{\citenamefont {Pe\~na}\ \emph {et~al.}(2005)\citenamefont {Pe\~na},
  \citenamefont {Sefrioui}, \citenamefont {Arias}, \citenamefont {Leon},
  \citenamefont {Santamaria}, \citenamefont {Martinez}, \citenamefont
  {te~Velthuis},\ and\ \citenamefont {Hoffmann}}]{PenaPRL2005}%
  \BibitemOpen
  \bibfield  {author} {\bibinfo {author} {\bibfnamefont {V.}~\bibnamefont
  {Pe\~na}}, \bibinfo {author} {\bibfnamefont {Z.}~\bibnamefont {Sefrioui}},
  \bibinfo {author} {\bibfnamefont {D.}~\bibnamefont {Arias}}, \bibinfo
  {author} {\bibfnamefont {C.}~\bibnamefont {Leon}}, \bibinfo {author}
  {\bibfnamefont {J.}~\bibnamefont {Santamaria}}, \bibinfo {author}
  {\bibfnamefont {J.~L.}\ \bibnamefont {Martinez}}, \bibinfo {author}
  {\bibfnamefont {S.~G.~E.}\ \bibnamefont {te~Velthuis}}, \ and\ \bibinfo
  {author} {\bibfnamefont {A.}~\bibnamefont {Hoffmann}},\ }\href {\doibase
  10.1103/PhysRevLett.94.057002} {\bibfield  {journal} {\bibinfo  {journal}
  {Phys. Rev. Lett.}\ }\textbf {\bibinfo {volume} {94}},\ \bibinfo {pages}
  {057002} (\bibinfo {year} {2005})}\BibitemShut {NoStop}%
\bibitem [{\citenamefont {Tagirov}(1999)}]{TagirovPRL1999}%
  \BibitemOpen
  \bibfield  {author} {\bibinfo {author} {\bibfnamefont {L.~R.}\ \bibnamefont
  {Tagirov}},\ }\href {\doibase 10.1103/PhysRevLett.83.2058} {\bibfield
  {journal} {\bibinfo  {journal} {Phys. Rev. Lett.}\ }\textbf {\bibinfo
  {volume} {83}},\ \bibinfo {pages} {2058} (\bibinfo {year}
  {1999})}\BibitemShut {NoStop}%
\bibitem [{\citenamefont {Buzdin}\ \emph {et~al.}(1999)\citenamefont {Buzdin},
  \citenamefont {Vedyayev},\ and\ \citenamefont {Ryzhanova}}]{BuzdinEPL1999}%
  \BibitemOpen
  \bibfield  {author} {\bibinfo {author} {\bibfnamefont {A.}~\bibnamefont
  {Buzdin}}, \bibinfo {author} {\bibfnamefont {A.}~\bibnamefont {Vedyayev}}, \
  and\ \bibinfo {author} {\bibfnamefont {N.}~\bibnamefont {Ryzhanova}},\ }\href
  {\doibase 10.1209/epl/i1999-00539-0} {\bibfield  {journal} {\bibinfo
  {journal} {Europhys. Lett.}\ }\textbf {\bibinfo {volume} {48}},\ \bibinfo
  {pages} {686} (\bibinfo {year} {1999})}\BibitemShut {NoStop}%
\bibitem [{\citenamefont {van Zalk}\ \emph {et~al.}(2009)\citenamefont {van
  Zalk}, \citenamefont {Veldhorst}, \citenamefont {Brinkman}, \citenamefont
  {Aarts},\ and\ \citenamefont {Hilgenkamp}}]{ZalkPRB2009}%
  \BibitemOpen
  \bibfield  {author} {\bibinfo {author} {\bibfnamefont {M.}~\bibnamefont {van
  Zalk}}, \bibinfo {author} {\bibfnamefont {M.}~\bibnamefont {Veldhorst}},
  \bibinfo {author} {\bibfnamefont {A.}~\bibnamefont {Brinkman}}, \bibinfo
  {author} {\bibfnamefont {J.}~\bibnamefont {Aarts}}, \ and\ \bibinfo {author}
  {\bibfnamefont {H.}~\bibnamefont {Hilgenkamp}},\ }\href {\doibase
  10.1103/PhysRevB.79.134509} {\bibfield  {journal} {\bibinfo  {journal} {Phys.
  Rev. B}\ }\textbf {\bibinfo {volume} {79}},\ \bibinfo {pages} {134509}
  (\bibinfo {year} {2009})}\BibitemShut {NoStop}%
\bibitem [{\citenamefont {Stamopoulos}\ \emph {et~al.}(2007)\citenamefont
  {Stamopoulos}, \citenamefont {Manios},\ and\ \citenamefont
  {Pissas}}]{StamopoulosSST2007}%
  \BibitemOpen
  \bibfield  {author} {\bibinfo {author} {\bibfnamefont {D.}~\bibnamefont
  {Stamopoulos}}, \bibinfo {author} {\bibfnamefont {E.}~\bibnamefont {Manios}},
  \ and\ \bibinfo {author} {\bibfnamefont {M.}~\bibnamefont {Pissas}},\ }\href
  {\doibase 10.1088/0953-2048/20/12/022} {\bibfield  {journal} {\bibinfo
  {journal} {Superconductor Science and Technology}\ }\textbf {\bibinfo
  {volume} {20}},\ \bibinfo {pages} {1205} (\bibinfo {year}
  {2007})}\BibitemShut {NoStop}%
\bibitem [{\citenamefont {Takahashi}\ \emph {et~al.}(1999)\citenamefont
  {Takahashi}, \citenamefont {Imamura},\ and\ \citenamefont
  {Maekawa}}]{TakahashPRL1999}%
  \BibitemOpen
  \bibfield  {author} {\bibinfo {author} {\bibfnamefont {S.}~\bibnamefont
  {Takahashi}}, \bibinfo {author} {\bibfnamefont {H.}~\bibnamefont {Imamura}},
  \ and\ \bibinfo {author} {\bibfnamefont {S.}~\bibnamefont {Maekawa}},\ }\href
  {\doibase 10.1103/PhysRevLett.82.3911} {\bibfield  {journal} {\bibinfo
  {journal} {Phys. Rev. Lett.}\ }\textbf {\bibinfo {volume} {82}},\ \bibinfo
  {pages} {3911} (\bibinfo {year} {1999})}\BibitemShut {NoStop}%
\bibitem [{\citenamefont {Nemes}\ \emph {et~al.}(2008)\citenamefont {Nemes},
  \citenamefont {Garc\'\i{}a-Hern\'andez}, \citenamefont {te~Velthuis},
  \citenamefont {Hoffmann}, \citenamefont {Visani}, \citenamefont
  {Garcia-Barriocanal}, \citenamefont {Pe\~na}, \citenamefont {Arias},
  \citenamefont {Sefrioui}, \citenamefont {Leon},\ and\ \citenamefont
  {Santamar\'\i{}a}}]{NemesPRB2008}%
  \BibitemOpen
  \bibfield  {author} {\bibinfo {author} {\bibfnamefont {N.~M.}\ \bibnamefont
  {Nemes}}, \bibinfo {author} {\bibfnamefont {M.}~\bibnamefont
  {Garc\'\i{}a-Hern\'andez}}, \bibinfo {author} {\bibfnamefont {S.~G.~E.}\
  \bibnamefont {te~Velthuis}}, \bibinfo {author} {\bibfnamefont
  {A.}~\bibnamefont {Hoffmann}}, \bibinfo {author} {\bibfnamefont
  {C.}~\bibnamefont {Visani}}, \bibinfo {author} {\bibfnamefont
  {J.}~\bibnamefont {Garcia-Barriocanal}}, \bibinfo {author} {\bibfnamefont
  {V.}~\bibnamefont {Pe\~na}}, \bibinfo {author} {\bibfnamefont
  {D.}~\bibnamefont {Arias}}, \bibinfo {author} {\bibfnamefont
  {Z.}~\bibnamefont {Sefrioui}}, \bibinfo {author} {\bibfnamefont
  {C.}~\bibnamefont {Leon}}, \ and\ \bibinfo {author} {\bibfnamefont
  {J.}~\bibnamefont {Santamar\'\i{}a}},\ }\href {\doibase
  10.1103/PhysRevB.78.094515} {\bibfield  {journal} {\bibinfo  {journal} {Phys.
  Rev. B}\ }\textbf {\bibinfo {volume} {78}},\ \bibinfo {pages} {094515}
  (\bibinfo {year} {2008})}\BibitemShut {NoStop}%
\bibitem [{\citenamefont {Dybko}\ \emph {et~al.}(2009)\citenamefont {Dybko},
  \citenamefont {Werner-Malento}, \citenamefont {Aleshkevych}, \citenamefont
  {Wojcik}, \citenamefont {Sawicki},\ and\ \citenamefont
  {Przyslupski}}]{DybkoPRB2009}%
  \BibitemOpen
  \bibfield  {author} {\bibinfo {author} {\bibfnamefont {K.}~\bibnamefont
  {Dybko}}, \bibinfo {author} {\bibfnamefont {K.}~\bibnamefont
  {Werner-Malento}}, \bibinfo {author} {\bibfnamefont {P.}~\bibnamefont
  {Aleshkevych}}, \bibinfo {author} {\bibfnamefont {M.}~\bibnamefont {Wojcik}},
  \bibinfo {author} {\bibfnamefont {M.}~\bibnamefont {Sawicki}}, \ and\
  \bibinfo {author} {\bibfnamefont {P.}~\bibnamefont {Przyslupski}},\ }\href
  {\doibase 10.1103/PhysRevB.80.144504} {\bibfield  {journal} {\bibinfo
  {journal} {Phys. Rev. B}\ }\textbf {\bibinfo {volume} {80}},\ \bibinfo
  {pages} {144504} (\bibinfo {year} {2009})}\BibitemShut {NoStop}%
\bibitem [{\citenamefont {Zhu}\ \emph {et~al.}(2010)\citenamefont {Zhu},
  \citenamefont {Krivorotov}, \citenamefont {Halterman},\ and\ \citenamefont
  {Valls}}]{ZhuPRL2010}%
  \BibitemOpen
  \bibfield  {author} {\bibinfo {author} {\bibfnamefont {J.}~\bibnamefont
  {Zhu}}, \bibinfo {author} {\bibfnamefont {I.~N.}\ \bibnamefont {Krivorotov}},
  \bibinfo {author} {\bibfnamefont {K.}~\bibnamefont {Halterman}}, \ and\
  \bibinfo {author} {\bibfnamefont {O.~T.}\ \bibnamefont {Valls}},\ }\href
  {\doibase 10.1103/PhysRevLett.105.207002} {\bibfield  {journal} {\bibinfo
  {journal} {Phys. Rev. Lett.}\ }\textbf {\bibinfo {volume} {105}},\ \bibinfo
  {pages} {207002} (\bibinfo {year} {2010})},\ \bibinfo {note} {the
  calculations are based on CuNi/Nb/CuNi trilayers.}\BibitemShut {Stop}%
\bibitem [{\citenamefont {Jaccarino}\ and\ \citenamefont
  {Peter}(1962)}]{JaccarinoPRL1962}%
  \BibitemOpen
  \bibfield  {author} {\bibinfo {author} {\bibfnamefont {V.}~\bibnamefont
  {Jaccarino}}\ and\ \bibinfo {author} {\bibfnamefont {M.}~\bibnamefont
  {Peter}},\ }\href {\doibase 10.1103/PhysRevLett.9.290} {\bibfield  {journal}
  {\bibinfo  {journal} {Phys. Rev. Lett.}\ }\textbf {\bibinfo {volume} {9}},\
  \bibinfo {pages} {290} (\bibinfo {year} {1962})}\BibitemShut {NoStop}%
\bibitem [{\citenamefont {Meul}\ \emph {et~al.}(1984)\citenamefont {Meul},
  \citenamefont {Rossel}, \citenamefont {Decroux}, \citenamefont {Fischer},
  \citenamefont {Remenyi},\ and\ \citenamefont {Briggs}}]{MeulPRL1984}%
  \BibitemOpen
  \bibfield  {author} {\bibinfo {author} {\bibfnamefont {H.~W.}\ \bibnamefont
  {Meul}}, \bibinfo {author} {\bibfnamefont {C.}~\bibnamefont {Rossel}},
  \bibinfo {author} {\bibfnamefont {M.}~\bibnamefont {Decroux}}, \bibinfo
  {author} {\bibfnamefont {O.}~\bibnamefont {Fischer}}, \bibinfo {author}
  {\bibfnamefont {G.}~\bibnamefont {Remenyi}}, \ and\ \bibinfo {author}
  {\bibfnamefont {A.}~\bibnamefont {Briggs}},\ }\href {\doibase
  10.1103/PhysRevLett.53.497} {\bibfield  {journal} {\bibinfo  {journal} {Phys.
  Rev. Lett.}\ }\textbf {\bibinfo {volume} {53}},\ \bibinfo {pages} {497}
  (\bibinfo {year} {1984})}\BibitemShut {NoStop}%
\bibitem [{\citenamefont {Felcher}\ \emph {et~al.}(1987)\citenamefont
  {Felcher}, \citenamefont {Hilleke}, \citenamefont {Crawford}, \citenamefont
  {Haumann}, \citenamefont {Kleb},\ and\ \citenamefont
  {Ostrowski}}]{FelcherRSI1987}%
  \BibitemOpen
  \bibfield  {author} {\bibinfo {author} {\bibfnamefont {G.~P.}\ \bibnamefont
  {Felcher}}, \bibinfo {author} {\bibfnamefont {R.~O.}\ \bibnamefont
  {Hilleke}}, \bibinfo {author} {\bibfnamefont {R.~K.}\ \bibnamefont
  {Crawford}}, \bibinfo {author} {\bibfnamefont {J.}~\bibnamefont {Haumann}},
  \bibinfo {author} {\bibfnamefont {R.}~\bibnamefont {Kleb}}, \ and\ \bibinfo
  {author} {\bibfnamefont {G.}~\bibnamefont {Ostrowski}},\ }\href {\doibase
  10.1063/1.1139225} {\bibfield  {journal} {\bibinfo  {journal} {Rev. Sci.
  Instrum.}\ }\textbf {\bibinfo {volume} {58}},\ \bibinfo {pages} {609}
  (\bibinfo {year} {1987})}\BibitemShut {NoStop}%
\bibitem [{\citenamefont {Majkrzak}(1996)}]{MajkrzakPhysB1996}%
  \BibitemOpen
  \bibfield  {author} {\bibinfo {author} {\bibfnamefont {C.}~\bibnamefont
  {Majkrzak}},\ }\href@noop {} {\bibfield  {journal} {\bibinfo  {journal}
  {Physica B: Condensed Matter}\ }\textbf {\bibinfo {volume} {221}},\ \bibinfo
  {pages} {342} (\bibinfo {year} {1996})}\BibitemShut {NoStop}%
\bibitem [{\citenamefont {Fitzsimmons}\ and\ \citenamefont
  {Majkrzak}(2005)}]{Fitz2005}%
  \BibitemOpen
  \bibfield  {author} {\bibinfo {author} {\bibfnamefont {M.~R.}\ \bibnamefont
  {Fitzsimmons}}\ and\ \bibinfo {author} {\bibfnamefont {C.}~\bibnamefont
  {Majkrzak}},\ }in\ \href {\doibase 10.1007/b101202} {\emph {\bibinfo
  {booktitle} {Modern Techniques for Characterizing Magnetic Materials}}},\
  \bibinfo {editor} {edited by\ \bibinfo {editor} {\bibfnamefont
  {Y.}~\bibnamefont {Zhu}}}\ (\bibinfo  {publisher} {Springer},\ \bibinfo
  {address} {US},\ \bibinfo {year} {2005})\ pp.\ \bibinfo {pages} {107 --
  155}\BibitemShut {NoStop}%
\bibitem [{\citenamefont {Sefrioui}\ \emph {et~al.}(2003)\citenamefont
  {Sefrioui}, \citenamefont {Arias}, \citenamefont {Pe\~na}, \citenamefont
  {Villegas}, \citenamefont {Varela}, \citenamefont {Prieto}, \citenamefont
  {Le\'on}, \citenamefont {Martinez},\ and\ \citenamefont
  {Santamaria}}]{SefriouiPRB2003}%
  \BibitemOpen
  \bibfield  {author} {\bibinfo {author} {\bibfnamefont {Z.}~\bibnamefont
  {Sefrioui}}, \bibinfo {author} {\bibfnamefont {D.}~\bibnamefont {Arias}},
  \bibinfo {author} {\bibfnamefont {V.}~\bibnamefont {Pe\~na}}, \bibinfo
  {author} {\bibfnamefont {J.~E.}\ \bibnamefont {Villegas}}, \bibinfo {author}
  {\bibfnamefont {M.}~\bibnamefont {Varela}}, \bibinfo {author} {\bibfnamefont
  {P.}~\bibnamefont {Prieto}}, \bibinfo {author} {\bibfnamefont
  {C.}~\bibnamefont {Le\'on}}, \bibinfo {author} {\bibfnamefont {J.~L.}\
  \bibnamefont {Martinez}}, \ and\ \bibinfo {author} {\bibfnamefont
  {J.}~\bibnamefont {Santamaria}},\ }\href {\doibase
  10.1103/PhysRevB.67.214511} {\bibfield  {journal} {\bibinfo  {journal} {Phys.
  Rev. B}\ }\textbf {\bibinfo {volume} {67}},\ \bibinfo {pages} {214511}
  (\bibinfo {year} {2003})}\BibitemShut {NoStop}%
\bibitem [{\citenamefont {Hoffmann}\ \emph {et~al.}(2005)\citenamefont
  {Hoffmann}, \citenamefont {te~Velthuis}, \citenamefont {Sefrioui},
  \citenamefont {Santamar\'\i{}a}, \citenamefont {Fitzsimmons}, \citenamefont
  {Park},\ and\ \citenamefont {Varela}}]{HoffmanPRB2005}%
  \BibitemOpen
  \bibfield  {author} {\bibinfo {author} {\bibfnamefont {A.}~\bibnamefont
  {Hoffmann}}, \bibinfo {author} {\bibfnamefont {S.~G.~E.}\ \bibnamefont
  {te~Velthuis}}, \bibinfo {author} {\bibfnamefont {Z.}~\bibnamefont
  {Sefrioui}}, \bibinfo {author} {\bibfnamefont {J.}~\bibnamefont
  {Santamar\'\i{}a}}, \bibinfo {author} {\bibfnamefont {M.~R.}\ \bibnamefont
  {Fitzsimmons}}, \bibinfo {author} {\bibfnamefont {S.}~\bibnamefont {Park}}, \
  and\ \bibinfo {author} {\bibfnamefont {M.}~\bibnamefont {Varela}},\ }\href
  {\doibase 10.1103/PhysRevB.72.140407} {\bibfield  {journal} {\bibinfo
  {journal} {Phys. Rev. B}\ }\textbf {\bibinfo {volume} {72}},\ \bibinfo
  {pages} {140407} (\bibinfo {year} {2005})}\BibitemShut {NoStop}%
\bibitem [{SM()}]{SM}%
  \BibitemOpen
  \href@noop {} {}\bibinfo {note} {See Supplemental Material.}\BibitemShut
  {Stop}%
\bibitem [{\citenamefont {Worthington}\ \emph {et~al.}(1987)\citenamefont
  {Worthington}, \citenamefont {Gallagher},\ and\ \citenamefont
  {Dinger}}]{WorthingtonPRL1987}%
  \BibitemOpen
  \bibfield  {author} {\bibinfo {author} {\bibfnamefont {T.~K.}\ \bibnamefont
  {Worthington}}, \bibinfo {author} {\bibfnamefont {W.~J.}\ \bibnamefont
  {Gallagher}}, \ and\ \bibinfo {author} {\bibfnamefont {T.~R.}\ \bibnamefont
  {Dinger}},\ }\href {\doibase 10.1103/PhysRevLett.59.1160} {\bibfield
  {journal} {\bibinfo  {journal} {Phys. Rev. Lett.}\ }\textbf {\bibinfo
  {volume} {59}},\ \bibinfo {pages} {1160} (\bibinfo {year}
  {1987})}\BibitemShut {NoStop}%
\bibitem [{\citenamefont {Li}\ \emph {et~al.}(1990)\citenamefont {Li},
  \citenamefont {Xi}, \citenamefont {Wu}, \citenamefont {Inam}, \citenamefont
  {Vadlamannati}, \citenamefont {McLean}, \citenamefont {Venkatesan},
  \citenamefont {Ramesh}, \citenamefont {Hwang}, \citenamefont {Martinez},\
  and\ \citenamefont {Nazar}}]{LiPRL1990}%
  \BibitemOpen
  \bibfield  {author} {\bibinfo {author} {\bibfnamefont {Q.}~\bibnamefont
  {Li}}, \bibinfo {author} {\bibfnamefont {X.~X.}\ \bibnamefont {Xi}}, \bibinfo
  {author} {\bibfnamefont {X.~D.}\ \bibnamefont {Wu}}, \bibinfo {author}
  {\bibfnamefont {A.}~\bibnamefont {Inam}}, \bibinfo {author} {\bibfnamefont
  {S.}~\bibnamefont {Vadlamannati}}, \bibinfo {author} {\bibfnamefont {W.~L.}\
  \bibnamefont {McLean}}, \bibinfo {author} {\bibfnamefont {T.}~\bibnamefont
  {Venkatesan}}, \bibinfo {author} {\bibfnamefont {R.}~\bibnamefont {Ramesh}},
  \bibinfo {author} {\bibfnamefont {D.~M.}\ \bibnamefont {Hwang}}, \bibinfo
  {author} {\bibfnamefont {J.~A.}\ \bibnamefont {Martinez}}, \ and\ \bibinfo
  {author} {\bibfnamefont {L.}~\bibnamefont {Nazar}},\ }\href {\doibase
  10.1103/PhysRevLett.64.3086} {\bibfield  {journal} {\bibinfo  {journal}
  {Phys. Rev. Lett.}\ }\textbf {\bibinfo {volume} {64}},\ \bibinfo {pages}
  {3086} (\bibinfo {year} {1990})}\BibitemShut {NoStop}%
\bibitem [{\citenamefont {Nemes}\ \emph {et~al.}(2010)\citenamefont {Nemes},
  \citenamefont {Visani}, \citenamefont {Leon}, \citenamefont
  {Garcia-Hernandez}, \citenamefont {Simon}, \citenamefont {Feher},
  \citenamefont {te~Velthuis}, \citenamefont {Hoffmann},\ and\ \citenamefont
  {Santamaria}}]{NemesAPL2010}%
  \BibitemOpen
  \bibfield  {author} {\bibinfo {author} {\bibfnamefont {N.~M.}\ \bibnamefont
  {Nemes}}, \bibinfo {author} {\bibfnamefont {C.}~\bibnamefont {Visani}},
  \bibinfo {author} {\bibfnamefont {C.}~\bibnamefont {Leon}}, \bibinfo {author}
  {\bibfnamefont {M.}~\bibnamefont {Garcia-Hernandez}}, \bibinfo {author}
  {\bibfnamefont {F.}~\bibnamefont {Simon}}, \bibinfo {author} {\bibfnamefont
  {T.}~\bibnamefont {Feher}}, \bibinfo {author} {\bibfnamefont {S.~G.~E.}\
  \bibnamefont {te~Velthuis}}, \bibinfo {author} {\bibfnamefont
  {A.}~\bibnamefont {Hoffmann}}, \ and\ \bibinfo {author} {\bibfnamefont
  {J.}~\bibnamefont {Santamaria}},\ }\href {\doibase DOI:10.1063/1.3464960}
  {\bibfield  {journal} {\bibinfo  {journal} {Appl. Phys. Lett.}\ }\textbf
  {\bibinfo {volume} {97}},\ \bibinfo {pages} {032501} (\bibinfo {year}
  {2010})}\BibitemShut {NoStop}%
\bibitem [{\citenamefont {Zhu}\ \emph {et~al.}(2009)\citenamefont {Zhu},
  \citenamefont {Cheng}, \citenamefont {Boone},\ and\ \citenamefont
  {Krivorotov}}]{ZhuPRL2009}%
  \BibitemOpen
  \bibfield  {author} {\bibinfo {author} {\bibfnamefont {J.}~\bibnamefont
  {Zhu}}, \bibinfo {author} {\bibfnamefont {X.}~\bibnamefont {Cheng}}, \bibinfo
  {author} {\bibfnamefont {C.}~\bibnamefont {Boone}}, \ and\ \bibinfo {author}
  {\bibfnamefont {I.~N.}\ \bibnamefont {Krivorotov}},\ }\href {\doibase
  10.1103/PhysRevLett.103.027004} {\bibfield  {journal} {\bibinfo  {journal}
  {Phys. Rev. Lett.}\ }\textbf {\bibinfo {volume} {103}},\ \bibinfo {pages}
  {027004} (\bibinfo {year} {2009})}\BibitemShut {NoStop}%
\bibitem [{\citenamefont {Parratt}(1954)}]{ParrattPR1954}%
  \BibitemOpen
  \bibfield  {author} {\bibinfo {author} {\bibfnamefont {L.~G.}\ \bibnamefont
  {Parratt}},\ }\href {\doibase 10.1103/PhysRev.95.359} {\bibfield  {journal}
  {\bibinfo  {journal} {Phys. Rev.}\ }\textbf {\bibinfo {volume} {95}},\
  \bibinfo {pages} {359} (\bibinfo {year} {1954})}\BibitemShut {NoStop}%
\bibitem [{\citenamefont {Stahn}\ \emph {et~al.}(2005)\citenamefont {Stahn},
  \citenamefont {Chakhalian}, \citenamefont {Niedermayer}, \citenamefont
  {Hoppler}, \citenamefont {Gutberlet}, \citenamefont {Voigt}, \citenamefont
  {Treubel}, \citenamefont {Habermeier}, \citenamefont {Cristiani},
  \citenamefont {Keimer},\ and\ \citenamefont {Bernhard}}]{StahnPRB2005}%
  \BibitemOpen
  \bibfield  {author} {\bibinfo {author} {\bibfnamefont {J.}~\bibnamefont
  {Stahn}}, \bibinfo {author} {\bibfnamefont {J.}~\bibnamefont {Chakhalian}},
  \bibinfo {author} {\bibfnamefont {C.}~\bibnamefont {Niedermayer}}, \bibinfo
  {author} {\bibfnamefont {J.}~\bibnamefont {Hoppler}}, \bibinfo {author}
  {\bibfnamefont {T.}~\bibnamefont {Gutberlet}}, \bibinfo {author}
  {\bibfnamefont {J.}~\bibnamefont {Voigt}}, \bibinfo {author} {\bibfnamefont
  {F.}~\bibnamefont {Treubel}}, \bibinfo {author} {\bibfnamefont {H.-U.}\
  \bibnamefont {Habermeier}}, \bibinfo {author} {\bibfnamefont
  {G.}~\bibnamefont {Cristiani}}, \bibinfo {author} {\bibfnamefont
  {B.}~\bibnamefont {Keimer}}, \ and\ \bibinfo {author} {\bibfnamefont
  {C.}~\bibnamefont {Bernhard}},\ }\href {\doibase 10.1103/PhysRevB.71.140509}
  {\bibfield  {journal} {\bibinfo  {journal} {Phys. Rev. B}\ }\textbf {\bibinfo
  {volume} {71}},\ \bibinfo {pages} {140509} (\bibinfo {year}
  {2005})}\BibitemShut {NoStop}%
\bibitem [{\citenamefont {Luo}\ \emph {et~al.}(2008)\citenamefont {Luo},
  \citenamefont {Pennycook},\ and\ \citenamefont {Pantelides}}]{LuoPRL2008}%
  \BibitemOpen
  \bibfield  {author} {\bibinfo {author} {\bibfnamefont {W.}~\bibnamefont
  {Luo}}, \bibinfo {author} {\bibfnamefont {S.~J.}\ \bibnamefont {Pennycook}},
  \ and\ \bibinfo {author} {\bibfnamefont {S.~T.}\ \bibnamefont {Pantelides}},\
  }\href {\doibase 10.1103/PhysRevLett.101.247204} {\bibfield  {journal}
  {\bibinfo  {journal} {Phys. Rev. Lett.}\ }\textbf {\bibinfo {volume} {101}},\
  \bibinfo {pages} {247204} (\bibinfo {year} {2008})}\BibitemShut {NoStop}%
\bibitem [{bkg()}]{bkg}%
  \BibitemOpen
  \href@noop {} {}\bibinfo {note} {There is an apparent difference in the PNR
  data between Figs.~\ref{Fig:InvM}a and ~\ref{Fig:DeadM}a because different
  background levels have been subtracted after optimizing the two
  models.}\BibitemShut {Stop}%
\end{thebibliography}%
\end{document}